# Ramsey CPT Signal Generation with a Miniature Clock Bench and a Dual-Frequency Optical Generator

J. Cotxet, F. Gutty, G. Baili, L. Morvan, D. Dolfi
Thales Research & Technology France
Palaiseau, France
jeremie.cotxet@thalesgroup.com

S. Guérandel, D. Holleville
LNE-SYRTE, Observatoire de Paris, UMR 8630
Paris, France

*Summary*—We demonstrate, for the first time, Ramsey CPT spectroscopy with a miniature electro-optical bench associated to a dual-frequency generator based on combined optical injection locking and optical phase locking techniques Preliminary results show Ramsey CPT resonance with a contrast of 2% at the Cesium D2 line. Frequency difference locking loops lead to a contribution of Dick effect to fractional frequency stability lower than $1.7 \times 10^{-13}$ at 1 s, in line with targeted clock stability of $5 \times 10^{-13}$ at 1 s.

*Keywords—compact clock; cesium; CPT; Ramsey interrogation.*

## I. Introduction

During the last few decades, new concepts and technologies have emerged, particularly in developing precision frequency standards using atomic vapor cells [1]. The concerted effort in this development is to improve the clock characteristic to achieve high frequency stability, and precision along with reduced volume, weight and power consumption. Among these technologies, coherent population trapping (CPT) in atomic vapor cells shows impressive frequency stability at both short and long-time time scales [2]. In this context, we propose to implement a highly compact CPT clock based on a miniature electro-optical bench and a dual-frequency optical generator based on optical injection locking technique for Lin per Lin interrogation [3]. The targeted relative frequency stability at 1 s of integration time is below $5 \times 10^{-13}$. Extensive studies of the miniature optical bench has been already reported for both short and long time scales and interactions with a dual-frequency beam were previously described [4]. Here we first present a dual-frequency optical generator based on the combination of two locking techniques: a passive optical injection-locking loop (OIL) and an optical phase-locking loop (OPLL). Then, based on the obtained dual-frequency generator, we present preliminary characterizations of continuous and pulsed CPT spectroscopy using the miniature clock bench.

## II. Experimental set-up

The first building block is the dual-frequency optical generator (see Fig.1). A low power DFB laser (master) is used to injection lock a low power DFB laser (slave) and generate the dual-frequency dual-polarization beams for our experiments. First, a part of the output from the master laser is sent into a fibred Mach-Zhender electro-optical intensity modulator (EOM) which is modulated at 9.2 GHz, i.e. the hyperfine frequency of cesium, delivered by a reference local oscillator (LO). The EOM is biased with a DC voltage set near to the carrier suppression DC bias. Then, the resulting optical field is fed back to the slave laser cavity with an injection ratio, r, defined as the ratio between the injected power and the slave laser emitted power. In order to take advantage of the optical injection loop (OIL) expected phase noise reduction, the optical delay between the master laser and the slave laser is compensated. To further stabilize the relative phase between the two lasers, an OPLL is implemented A photodiode (FPD) and an amplifier (RFA) are used to detect the beat note at a part of the output of the injection-locked diode laser system. The corresponding signal is compared with a microwave mixer to the 9.2 GHz signal coming from a reference local oscillator. The resulting error signal is filtered and fed back to slave laser, through its current supply. The resulting CPT beams are cross polarized and sent to the clock bench.

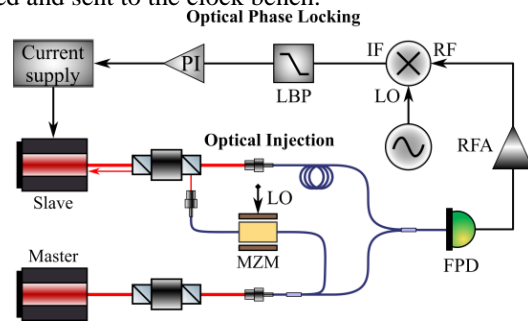

Fig. 1. General scheme of the combined loops OIL and OPLL.

The second building block is the miniature electro-optical bench (see Fig2). It provides all the functions needed for the CPT beam stabilization within a low volume of 10 L. The electro-optical bench functions description and validation are detailed in [4]. The cross-polarized laser beams are superimposed and sent through the miniature electro-optical bench. At the last stage, the optical beam, with a mean power of 1mW per polarization, is extended and sent to a glass cell containing the cesium vapor and a buffer gas mixture. This cell is included in a physics package that ensures temperature and static magnetic environment stabilization. The transmitted light is detected with a photodiode, and the obtained CPT signal is processed.

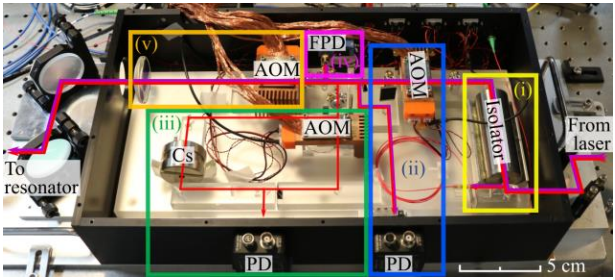

Fig. 2. Miniature electro-optical breadboard. The dimensions are 9.5 × 25 × 43 cm3. (i) optical isolation, (ii) Power stabilization (iii) optical frequency locking, (iv) beat note detection (v) Optical pulses shaping

## III. RESULTS

First, the dual-frequency generator is operating only with the OIL. As expected from this injection technique, we observe a drastic and a wideband reduction of the beat note phase noise, obtained for an injection ratio of -30 dB (See Fig.3).

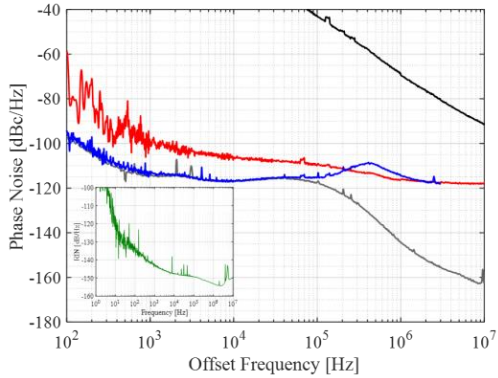

Fig. 3. Phase noise of the beat note between the two CPT beams. Black line: free-running, Red line: with OIL and -30dB injection ration, Blue line with OIL and OPLL. Grey line: phase noise of the LO at 9.2 GHz Inset: RIN spectrum of the CPT beams in free running.

From 100 Hz to 10 kHz, the phase noise (red line) reaches a low noise level, going from -60 dBc/Hz to -105 dBc/Hz. At 100 kHz, it represents an improvement of more than 70 dB compared with the free-running beat note phase noise (black line). When the OPLL is applied (blue line), the phase noise of the beat note reaches its lowest limit for frequencies going up to 200 kHz., i.e. the phase noise of the LO (grey line). Due to OPLL bandwidth limitation, 5 dB noise excess is measured around 400 kHz. The laser power stabilization loop is not applied, leading to the relative intensity noise (RIN) spectrum depicted by the insert of Fig. 3.

Zeeman split CPT resonances are first observed using continuous laser beams and after applying a DC axial magnetic field of 8 µT (see Fig. 4-left). The 0-0 CPT resonance is observed at the center with a linewidth of 2.4 kHz et and a contrast of 2.3%, 4 times lower than reported values with Cs D1 line [5]. Next, we used pulsed CPT beams to generate Ramsey fringes. A first pulse of length $\tau_p$ (2ms) allows to pump the atoms into the CPT state. Atoms then evolve freely in the dark during a time $T_R$. A second pulse is used for CPT signal detection. A delay $\tau_R$ of 20 µs is taken before opening a detection window with a length $\tau_D$ (25 µs).

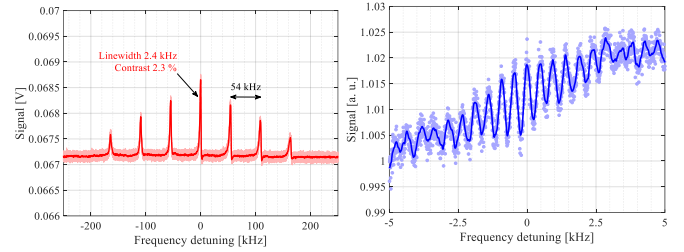

Fig. 4. Left: Zeeman split CPT resonances observed with CW optical beams. Right: Ramsey CPT resonances observed with pulsed optical beams

Fig. 4-right shows preliminary Ramsey-CPT fringes for a free-evolution time $T_R$ of 4 ms. The central fringe linewidth is about 125 Hz. The CPT contrast is about 2%. Considering the OPLL phase noise and the laser free-running RIN depicted in Fig.3, their contribution to the clock fractional frequency stability at 1 s integration, respectively through Dick effect and light shift effects, respectively, are equal to $1.7 \times 10^{-13}$ and $3.7 \times 10^{-13}$.

## IV. CONCLUSIONS

Ramsey CPT spectroscopy is performed with a miniature electro-optical bench (volume < 10 L) associated to a dual-frequency generator based on combined optical injection locking and optical phase locking techniques of two DFB COTS lasers. Preliminary results show Ramsey CPT resonance with a contrast of 2% at the Cesium D2 line. Frequency difference locking loops lead to a contribution of Dick effect to fractional frequency stability lower than $1.7 \times 10^{-13}$ at 1 s, in line with targeted clock stability of $5 \times 10^{-13}$ at 1 s. Further investigations should include the simultaneous operation of stabilization servo loops and an optimization of Ramsey fringes contrast.

This work was partially supported by the Direction Générale de l'Armement and was performed in the framework of J. Cotxet thesis supported by ANRT.